\documentclass[reprint,aps,twocolumn,longbibliography,superscriptaddress,amsmath,amssymb]{revtex4-2}

\usepackage{graphicx} 
\usepackage{enumitem}
\usepackage{array}
\usepackage{newtx}
\usepackage{fancyhdr}
\usepackage{parskip}
\usepackage{framed}
\usepackage{amsmath}
\usepackage{bm}
\usepackage{braket}
\usepackage{upgreek}
\usepackage{gensymb}
\usepackage{lipsum}
\usepackage{silence}
\usepackage{orcidlink}
\usepackage[normalem]{ulem}

\definecolor{myColor}{rgb}{0.02,0.12,0.3}

\def\be{\begin{equation}}
\def\ee{\end{equation}}

\makeatletter
\def\@fnsymbol#1{\ensuremath{\ifcase#1\or \dagger\or *\or \ddagger\or
   \mathsection\or \mathparagraph\or \|\or **\or \dagger\dagger
   \or \ddagger\ddagger \else\@ctrerr\fi}}
\makeatother

\begin{document} 

\title{
Tuning interactions between static-field-shielded polar molecules with microwaves
}
\author{Christopher J. Ho\orcidlink{0000-0002-8990-8159}}\email{christopher.ho15@imperial.ac.uk}
\affiliation{Centre for Cold Matter, Blackett Laboratory, Imperial College London, London SW7 2AZ, United Kingdom\vspace{-0.5em}}
\author{Joy Dutta\orcidlink{0000-0002-9345-257X}}
\affiliation{Joint Quantum Centre (JQC) Durham-Newcastle, Department of Chemistry, Durham University, South Road, Durham DH1 3LE, United Kingdom\vspace{-0.5em}}
\author{Bijit Mukherjee\orcidlink{0000-0001-9242-8378}}
\thanks{Present address: School of Chemical Sciences, National Institute of Science Education and Research (NISER), Bhubaneswar, An OCC of Homi Bhabha National Institute, Khurda, Odisha 752050, India}
\affiliation{Joint Quantum Centre (JQC) Durham-Newcastle, Department of Chemistry, Durham University, South Road, Durham DH1 3LE, United Kingdom\vspace{-0.5em}}
\affiliation{Faculty of Physics, University of Warsaw, Pasteura 5, 02-093 Warsaw, Poland\vspace{-0.5em}}
\author{Jeremy M. Hutson\orcidlink{0000-0002-4344-6622}}\email{j.m.hutson@durham.ac.uk}
\affiliation{Joint Quantum Centre (JQC) Durham-Newcastle, Department of Chemistry, Durham University, South Road, Durham DH1 3LE, United Kingdom\vspace{-0.5em}}
\author{Michael R. Tarbutt\orcidlink{0000-0003-2713-9531}}\email{m.tarbutt@imperial.ac.uk}
\affiliation{Centre for Cold Matter, Blackett Laboratory, Imperial College London, London SW7 2AZ, United Kingdom\vspace{-0.5em}}

\begin{abstract}
The ability to tune interparticle interactions is one of the main advantages of using ultracold quantum gases for quantum simulation of many-body physics. Current experiments with ultracold polar molecules employ shielding with microwave or static electric fields to prevent destructive collisional losses. The interaction potential of microwave-shielded molecules can be tuned by using microwaves of two different polarizations, while for static-field-shielded molecules the tunability of interactions is more limited and depends on the particular species. In this work, we propose a general method to tune the interactions between static-field-shielded molecules by applying a microwave field. We carry out coupled-channel scattering calculations in a field-dressed basis set to determine loss rate coefficients and scattering lengths.  We find that both the s-wave scattering length and the dipole length can be widely tuned by changing the parameters of the microwave field, while maintaining strong suppression of lossy collisions. 
\end{abstract}

\maketitle

\section{Introduction}

Ultracold gases of polar molecules have recently emerged as a promising platform for exploring strongly interacting dipolar quantum systems. While quantum phases unique to dipolar systems, such as droplets and supersolids, have been intensively studied using ultracold gases of highly magnetic atoms~\cite{Lahaye2009,Chomaz2022}, novel strongly correlated quantum phases beyond the mean-field regime are predicted to emerge in ultracold polar molecular gases owing to the much stronger dipole-dipole interactions between molecules~\cite{Langen2024, Schindewolf2025}. The ability to tune interparticle interactions while maintaining collisional stability of the ultracold gas~\cite{Karman2025,Yuan2025} is crucial for the exploration of such phases, including novel density-modulated structures~\cite{Schmidt2022} and self-assembled dipolar crystals~\cite{Astrakharchik2007,Buchler2007PRL,Ciardi2025}.

A long-standing challenge that stood in the way of achieving quantum-degenerate polar molecular gases was the near-universal two-body collisional losses observed across all molecular species~\cite{Bause2023}. Recently, this was overcome by using a static electric field~\cite{Avdeenkov2006, Wang2015, Matsuda2020} or a circularly polarized microwave field~\cite{Karman2018,Lassabliere2018,Anderegg2021} to engineer a long-range interaction potential that shields the molecules from short-range collisions.  Moreover, it was later demonstrated that, by adding a linearly polarized microwave field to the circularly polarized one, this interaction potential can be tuned over a wide range to realise different regimes of quantum matter~\cite{Karman2025,Yuan2025}. The successful implementation of these collisional shielding techniques led to the achievement of quantum-degenerate Fermi gases~\cite{Valtolina2020,Schindewolf2022} and Bose--Einstein condensates (BECs) of polar molecules~\cite{Bigagli2024,Shi2025}. For the creation of BECs, only microwave shielding has been used so far. Nevertheless, calculations show that, for many molecules of interest, static-field shielding offers a far greater suppression of loss~\cite{Mukherjee2023,Mukherjee2024}. Because of this, it is preferred during the long evaporative-cooling stage. The disadvantage is that, once quantum degeneracy is achieved, the static shield offers limited tunability of the interactions. Such tunability is essential for exploring novel phases and dynamics of molecular BECs, in the same way that Feshbach resonances have been crucial for ultracold atomic gases~\cite{Chin2010}.

In this work, we show that the interaction potential between static-field-shielded molecules can also be widely tuned by adding a circularly polarized microwave field. Combining static electric fields with microwaves for shielding has been considered in the past. B\"uchler \emph{et al.}~\cite{Buchler2007PRL, Buchler2007NP} studied the effective two- and three-body interaction potentials of dipolar molecules in combined static and microwave fields. Building on those ideas, Gorshkov \emph{et al.}~\cite{Gorshkov2008} proposed to cancel the first-order dipole-dipole interactions of the static and microwave fields through careful tuning of the field parameters, realising a completely repulsive shield from the second-order interactions, while Karman~\cite{Karman2020} considered supplementing an elliptically polarized microwave field with a small static field, such that the molecules are shielded by an effectively pure circularly polarized microwave field. Schmidt \emph{et al.}~\cite{Schmidt2022} considered microwave shielding with the addition of a small static field to tune the interactions. Our proposal differs from these earlier ideas in that the collisional shielding arises mostly from the static electric field~\cite{Mukherjee2023}; the role of the microwave field is to tune the interaction parameters by modifying the long-range interaction potential. We carry out our calculations for CaF, which is particularly well suited to static-field shielding~\cite{Mukherjee2023} as both two-body and three-body losses are strongly suppressed. We find that the characteristic length scales of the interaction, the s-wave scattering length and the dipole length, can be tuned in both sign and magnitude, and that collisional shielding remains effective throughout.

\section{Concept}

\begin{figure}[t]
\centerline{\includegraphics[width=\columnwidth]{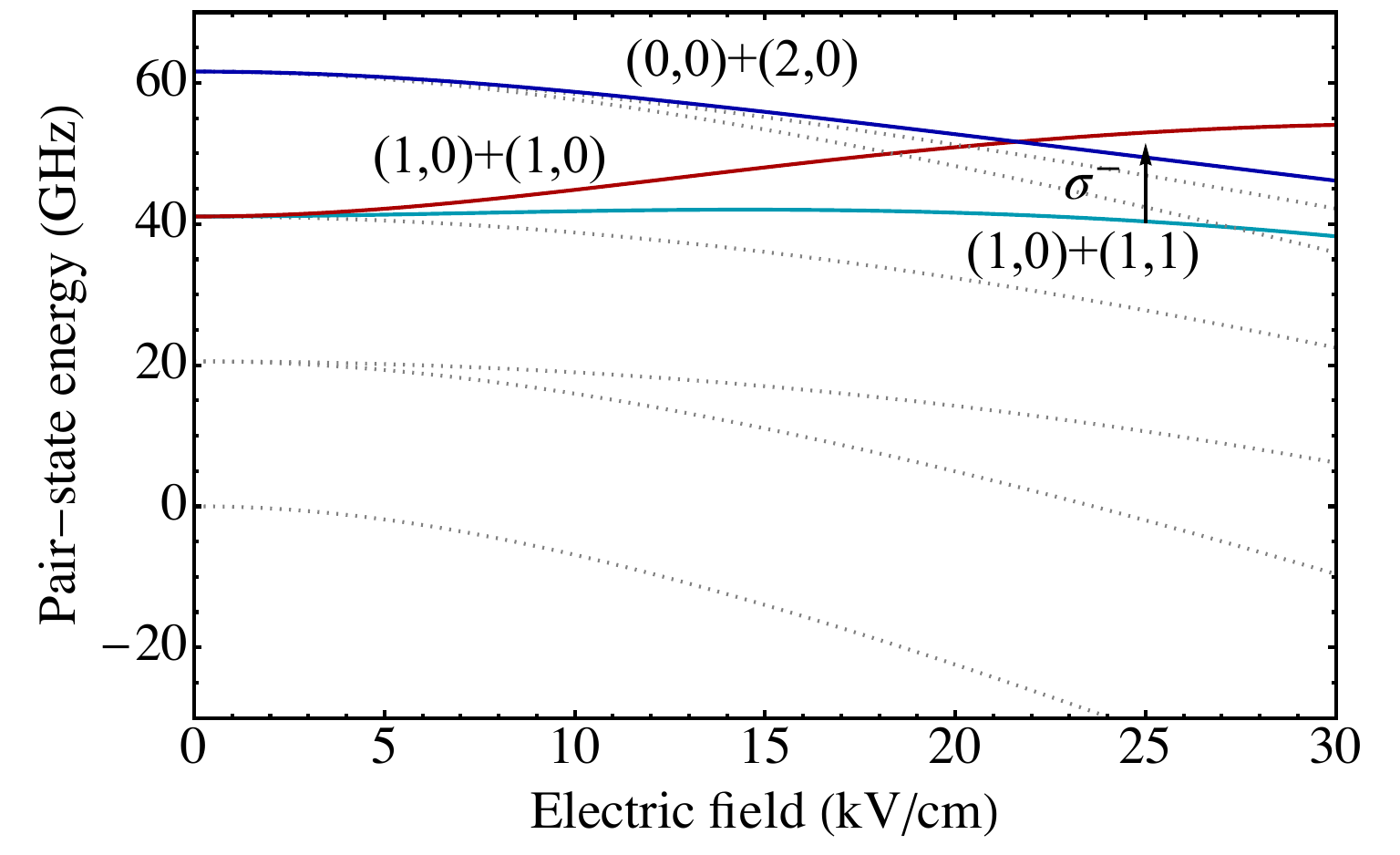}}
\caption{
Energy-level diagram for pairs of CaF molecules. The highlighted states are labeled by hindered-rotor quantum numbers $(\tilde{n}, \lvert m_n \rvert)$. We apply a near-resonant microwave field with circular polarization, which couples $(1,0)$ to $(1,1)$. 
}
\label{fig1}
\vspace{-0.5em}
\end{figure}

Static-field shielding uses an electric field just above the resonant field where two pair states, $(1,0)+(1,0)$ and $(0,0)+(2,0)$, cross in energy~\cite{Mukherjee2023}. This field is $F_{\rm X}=3.244b/\mu$, where $b$ is the molecular rotational constant and $\mu$ is its dipole moment. The crossing is shown for CaF in Fig.~\ref{fig1}, where $F_{\rm X}=21.55\,\mathrm{kV/cm}$. The molecular states are labeled by quantum numbers $(\tilde{n}, m_n)$: $\tilde{n}$ is a hindered-rotor quantum number that correlates adiabatically with the rotational angular momentum $n$ of the corresponding field-free state and $m_n$ is its conserved projection along the electric field.

To modify the long-range part of the interaction potential, we add a near-resonant, circularly polarized microwave field, as shown in Fig.~\ref{fig1}(a). In this case, we use red-detuned microwaves with $\sigma^-$ polarization to couple $(1,0)$ to $(1,1)$. We denote the Rabi frequency of the microwaves by $\Omega$ and define the detuning by $\Delta = \omega_0 - \omega$, where $\omega_0$ is the transition frequency and $\omega$ is the microwave frequency.  The microwave field gives rise to a dipole-dipole interaction that has the opposite sign to the one generated by the static electric field. The strength of the total dipole-dipole interaction can therefore be tuned using the microwave parameters, and can even be compensated entirely such that the remaining interaction is entirely repulsive. As the degree of microwave coupling is solely determined by the ratio $\Delta/\Omega$, for a fixed value of the electric field, there is a value of $\Delta/\Omega$ where the two opposite contributions to the dipole-dipole interaction cancel out and the effective dipole moment of the molecules is zero.

\section{Adiabatic potentials}

In a rigid-rotor approximation, the Hamiltonian for a single molecule $i$ in the presence of a static electric field $\boldsymbol{F}$ and a microwave field is
\begin{equation}
    \hat{h}_i = b\hat{\boldsymbol{n}}_i^2 - \boldsymbol{\mu}_i\cdot\boldsymbol{F} + \hat{h}_\mathrm{mw},
\label{eq1}
\end{equation}
where $\hat{\boldsymbol{n}}^2$ is the squared rotational angular momentum operator and $\boldsymbol{\mu}$ is the dipole moment vector.  In the absence of the microwave field, the eigenstates are $\ket{\tilde{n},m_n} = \sum_n a_n \ket{n, m_n}$ and are calculated in a well-converged basis set with $n_\textrm{max}=5$.  We add a $\sigma^-$-polarized microwave field which couples the states $(\tilde{n},m_n)=(1,0)$ and $(1,1)$ in the electric field, so that $\hat{h}_\mathrm{mw} = \hbar\Omega\cos\omega t\left[\ket{1,0}\bra{1,1}+\ket{1,1}\bra{1,0}\right]$.  For simplicity we consider only the state space spanned by $\{ \ket{1,0}, \ket{0,0}, \ket{2,0}, \ket{1,\pm1} \}$. After applying the rotating-wave approximation, the eigenstates of the Hamiltonian of Eq.~(\ref{eq1}) are $\ket{+} = ue^{-i\omega t}\ket{1,1}+v\ket{1,0}$, $\ket{-} = ve^{-i\omega t} \ket{1,1} - u\ket{1,0}$, $\ket{\rm D} = e^{-i\omega t}\ket{1,-1}$, $\ket{0} = \ket{0,0}$, and $\ket{2} = \ket{2,0}$, where $u=\sqrt{(1-\Delta/\Omega_\mathrm{eff})/2}$, $v=\sqrt{(1+\Delta/\Omega_\mathrm{eff})/2}$, and $\Omega_\mathrm{eff} = \sqrt{\Omega^2+\Delta^2}$ is the effective Rabi frequency. Here, $\ket{\pm}$ represent microwave-induced superpositions of static-field-dressed states, whereas $\ket{\rm D}$ represents a dark state that is not coupled by the microwave field but is important in collisions.

The Hamiltonian that describes the relative motion of two molecules is
\begin{equation}
\begin{aligned}
    \hat{H}&=\frac{\hbar^2}{2\mu_{\rm red}}\left(-\frac{1}{R}\frac{d^2}{dR^2}R + \frac{\hat{\boldsymbol{L}}^2}{R^2} \right) + \hat{h}_1 + \hat{h}_2 + \hat{V}_\mathrm{int} \\
    &= -\frac{\hbar^2}{2\mu_{\rm red} R}\frac{d^2}{dR^2}R + \hat{H}_\textrm{internal},
\end{aligned}
\end{equation}
where $\mu_{\rm red}$ is the reduced mass, $R$ is the separation between the molecules and $\hat{\boldsymbol{L}}^2$ is the relative angular momentum operator. $V_\mathrm{int}$ is the two-body interaction potential, which here is the dipole-dipole interaction, 
\begin{equation}
    \hat{V}_\mathrm{int} = \frac{ \boldsymbol{\mu}_1 \cdot \boldsymbol{\mu}_2 - 3(\boldsymbol{\mu}_1 \cdot \hat{\boldsymbol{R}})(\boldsymbol{\mu}_2 \cdot \hat{\boldsymbol{R}})}{4\pi\epsilon_0R^3},
\end{equation}
where $\hat{\boldsymbol{R}}$ is the unit vector along the line joining the two molecules.

We use the basis set $\ket{p}_\textrm{pair}\ket{L,M_L}$, where $\ket{L,M_L}$ represents an eigenstate of $\hat{\boldsymbol{L}}^2$ and $\ket{p}_\textrm{pair}$ represents a pair state constructed from the single-molecule ones defined above,
\begin{equation}
\begin{aligned}
\ket{1}_\textrm{pair} &= \ket{+} \ket{+} , \\
\ket{2}_\textrm{pair} &= \frac{1}{\sqrt{2}} \left( \ket{+} \ket{\rm D} + \ket{\rm D} \ket{+} \right), \\
\ket{3}_\textrm{pair} &= \frac{1}{\sqrt{2}} \left( \ket{+} \ket{-} + \ket{-} \ket{+} \right), \\
\ket{4}_\textrm{pair} &= \frac{1}{\sqrt{2}} \left( \ket{-} \ket{\rm D} + \ket{\rm D} \ket{-} \right), \\
\ket{5}_\textrm{pair} &= \ket{-} \ket{-}, \\
\ket{6}_\textrm{pair} &= \frac{1}{\sqrt{2}} \left( \ket{0} \ket{2} + \ket{2} \ket{0} \right).
\end{aligned}
\end{equation}
This basis set is chosen to include only states that are sufficiently close in energy, i.e. no further apart than the energy difference between the states used for static-field shielding. For bosonic (fermionic) molecules, $L$ must be even (odd).
In this basis set, $\hbar^2\hat{\boldsymbol{L}}^2/(2\mu_{\rm red} R^2)+\hat{h}_1 + \hat{h}_2$ has only diagonal matrix elements, $\hbar^2L(L+1)/(2\mu_{\rm red} R^2)+E_\mathrm{pair}$, where $E_\mathrm{pair}/\hbar = \{ \Omega_\mathrm{eff}-\Delta, (\Omega_\mathrm{eff}-3\Delta)/2, -\Delta, (-\Omega_\mathrm{eff}-3\Delta)/2,  -\Omega_\mathrm{eff}-\Delta, -\Delta_\mathrm{S} \}$ and $\hbar\Delta_\mathrm{S}$ is the Stark shift between $(1,0)+(1,0)$ and $(0,0)+(2,0)$ at the chosen electric field. $\hat{V}_\mathrm{int}$ has both diagonal and off-diagonal matrix elements in this basis. The adiabatic potentials $V_j(R)$ are the eigenvalues obtained by diagonalizing $\hat{H}_\textrm{internal}$ at fixed $R$.

\begin{figure}[t]
\centerline{\includegraphics[width=\columnwidth]{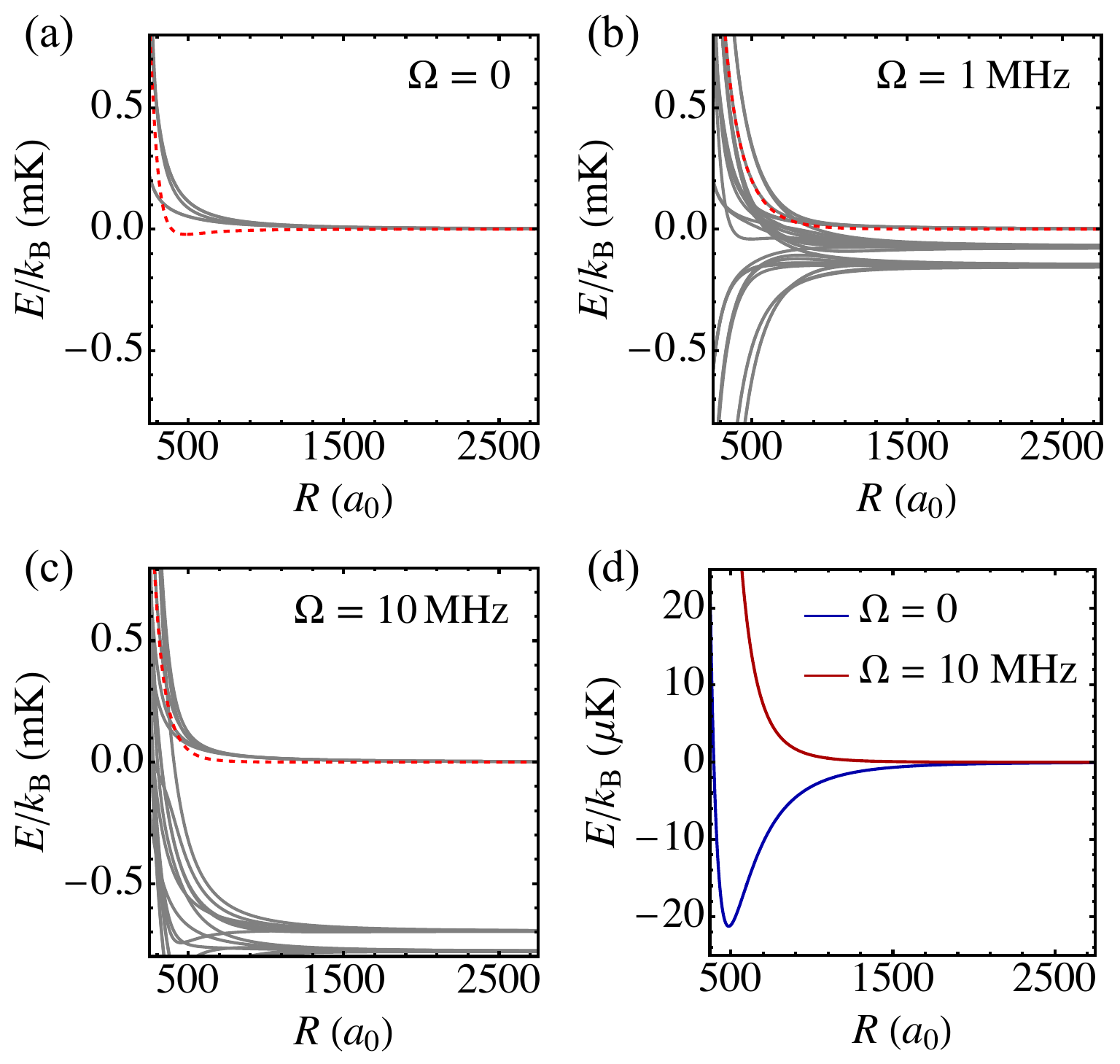}}
\caption{Adiabatic potential curves for two CaF molecules at a static electric field of $F=22.5\,\mathrm{kV/cm}$, showing curves only for $L=0$ and 2. (a) No microwave field ($\Omega=0$). The red dashed curve is the potential for s-wave scattering that correlates with $(1,0)$ + $(1,0)$. (b) A weak microwave field ($\Omega=1\,\mathrm{MHz}$, $\Delta/\Omega = 1.3$); other pair states dressed by the microwave field are now close in energy to the incoming state. (c) A stronger microwave field ($\Omega=10\,\mathrm{MHz}$, $\Delta/\Omega=1.3$); the energy separation between the incoming and microwave-dressed pair states is larger. (d) Expanded view of the highlighted curves in (a) and (c). The shallow potential well that exists for $\Omega=0$ disappears at the compensation point $\Delta/\Omega=1.3$.}
\label{fig2}
\vspace{-0.5em}
\end{figure}

Figure \ref{fig2} shows the resulting adiabatic potentials at an electric field of $F=22.5\,\mathrm{kV/cm}$. In Fig.~\ref{fig2}(a), no microwave field is present and we obtain the static-field-shielded potentials correlating with the incoming state $(1,0)$ + $(1,0)$. We are most interested in the potential for s-wave scattering, $L=0$, which is highlighted as a red dashed curve. The repulsive shield of this potential arises from dipole-dipole coupling to the state $(0,0)$ + $(2,0)$. 
In Fig.~\ref{fig2}(b), a microwave field with $\Omega=1\,\mathrm{MHz}$ and $\Delta/\Omega=1.3$ is added, which brings pair states 2 to 5 close in energy to states 1 and 6. These states couple to pair state 1 through dipole-dipole interactions, modifying the adiabatic potentials. Increasing $\Omega$ and $\Delta$, while keeping $\Delta/\Omega$ the same, separates these states further, as shown in Fig.~\ref{fig2}(c). For this particular electric field, the dipole-dipole interactions induced by the static and microwave fields cancel out at $\Delta/\Omega=1.3$, resulting in a completely repulsive potential. The same cancellation is achieved at the same value of $\Delta/\Omega$ for other polar molecules at the same value of $F/F_\textrm{X}$. In the region where static-field shielding is effective, the value of $\Delta/\Omega$ required depends only weakly on $F$. The effect is shown more clearly in Fig.~\ref{fig2}(d), where we directly compare the incoming s-wave adiabats for $\Omega=0$ and $\Omega=10\,\mathrm{MHz}$ at this compensation point. The shallow potential well present for pure static-field shielding completely disappears when the microwave field is switched on.

\section{Coupled-channel calculations}

Calculations on a single adiabat are valuable for understanding the long-range potential and the existence of a shielding barrier and potential well, but they do not take account of the nonadiabatic processes that are the main source of collisional loss. The lower adiabats are generally unshielded, so collisions that transfer to them usually result in loss of both particles, either by inelastic loss or via short-range processes such as chemical reaction or collision-induced absorption of photons from the trapping lasers. To take account of this, we have implemented full coupled-channel calculations in the presence of both static-electric and microwave fields as a plug-in basis-set suite for the MOLSCAT package \cite{molscat:2019, mbf-github:2023}. The methods are formally similar to those of Refs.\ \cite{Karman2018} and \cite{Dutta:universal:2025}, but replace the free-rotor functions used there with static-field-dressed functions as in Ref.\ \cite{Mukherjee2023}. We take account of short-range loss by imposing a fully absorbing boundary condition at short range. The calculations use a basis set that explicitly includes static-field-dressed pair functions corresponding to the states $(0,0)$, $(1,0)$, $(2,0)$ and $(1,\pm 1)$, with all partial waves up to $L_\textrm{max}=12$. Pair functions outside this space are included by a Van Vleck transformation as described in Ref.\ \cite{Mukherjee2023}. The calculations produce scattering matrices ${\bf S}$ that may be used to evaluate cross sections and rate coefficients for elastic scattering and loss as described in Ref.\ \cite{Mukherjee2023}. They also provide complex scattering lengths $a=\alpha-i\beta$, where the real part $\alpha$ dominates when shielding is effective.

\begin{figure}[t]
\centerline{\includegraphics[width=\columnwidth]{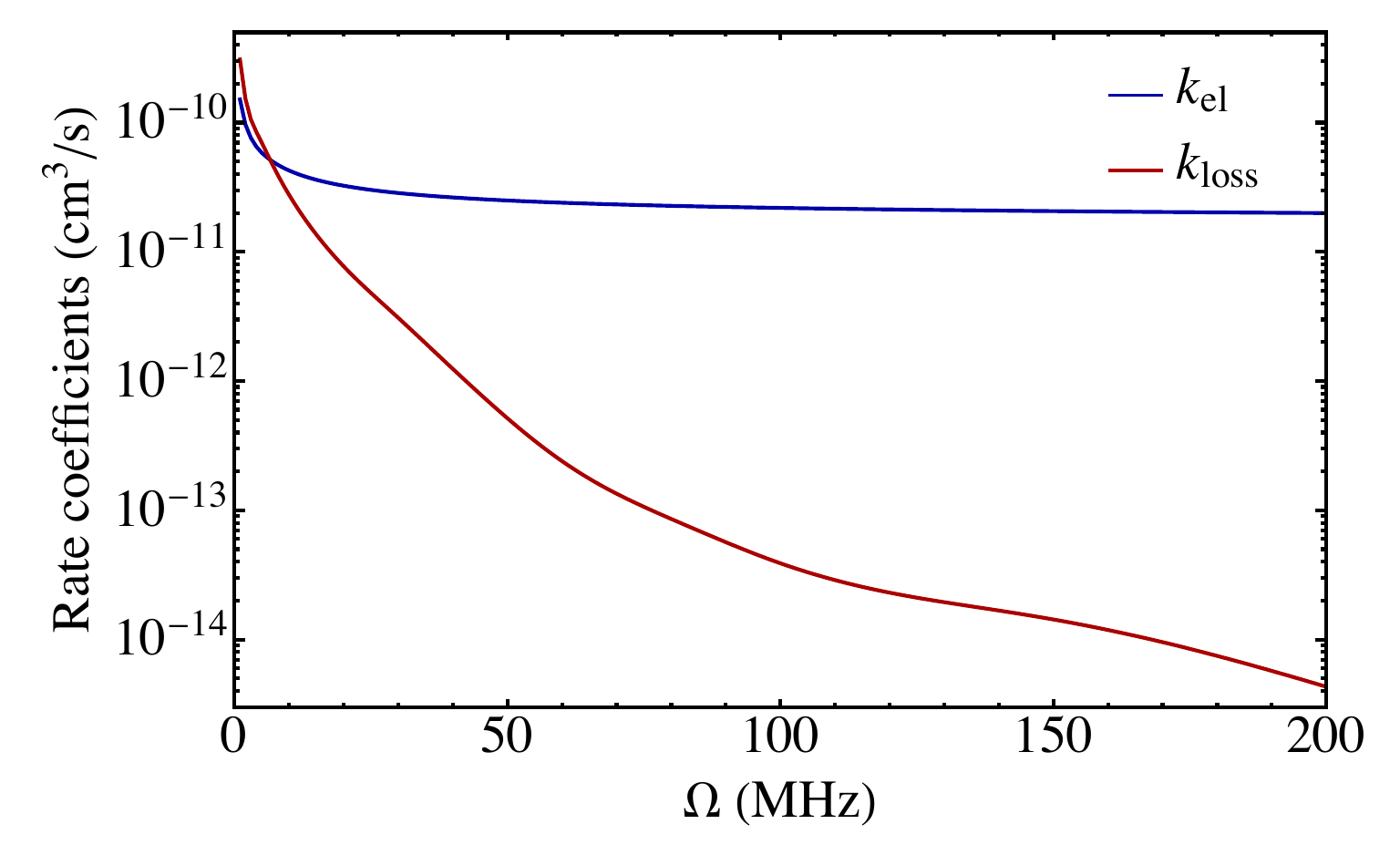}}
\caption{
Rate coefficients for elastic scattering and total loss, obtained from coupled-channel calculations at the compensation point $\Delta/\Omega = 1.3$, as a function of $\Omega$.
}
\label{fig:rate_vs_Omega}
\vspace{-0.5em}
\end{figure}

Figure \ref{fig:rate_vs_Omega} shows the rate coefficients for elastic scattering and total loss at the compensation point as a function of $\Omega$, calculated at a collision energy of 10~nK~$\times\ k_{\rm B}$. At low values of $\Omega$ there is fast loss. This is due to non-adiabatic transitions to the microwave-dressed pair states, which are unshielded and are in close proximity to the incoming state, as shown in Fig.~\ref{fig2}(b). Increasing $\Omega$ decreases this loss by separating the states further, as shown in Fig.~\ref{fig2}(c). At $\Omega=60\,\mathrm{MHz}$ and $F=22.5\,\mathrm{kV/cm}$, the rate coefficient for total loss is $2.4 \times 10^{-13}\,\mathrm{cm^3/s}$. For CaF molecules at this field, this Rabi frequency corresponds to a microwave intensity of $I=24\,\mathrm{W/cm^2}$. A typical molecular BEC has a density of $\sim 10^{12}\,\mathrm{cm^{-3}}$~\cite{Bigagli2024}, so with this loss rate the BEC will have a lifetime of several seconds.

\begin{figure}[t]
\centerline{\includegraphics[width=\columnwidth]{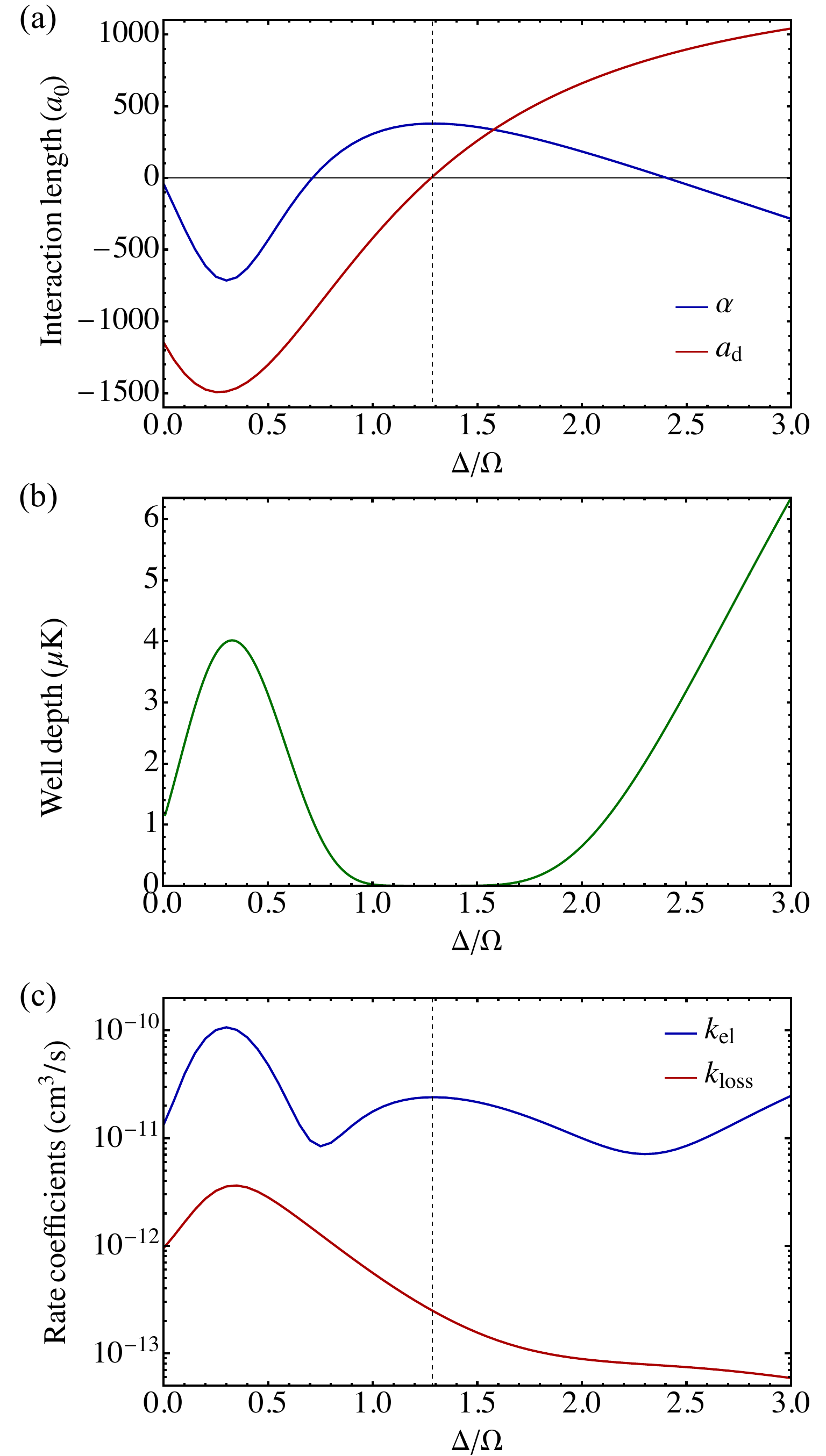}}
\caption{
Characteristics of the two-body system as a function of $\Delta/\Omega$ (a) The real part $\alpha$ of the s-wave scattering length, and the dipole length $a_\mathrm{d}$.
(b) Depth of the long-range potential well for the incoming state with $L=0$. 
(c) Rate coefficients for elastic scattering and total loss, obtained from coupled-channel calculations.
}
\label{fig:cc}
\vspace{-0.5em}
\end{figure}

Figure \ref{fig:cc}(a,b) show the interaction lengths and well depth as a function of $\Delta/\Omega$, all at $\Omega=60$~MHz.  Figure \ref{fig:cc}(a) shows the real part, $\alpha$, of the scattering length, and the dipole length defined as $a_\mathrm{d} = \mu_{\rm red} d_\mathrm{eff}^2/(4\pi\epsilon_0\hbar^2)$. Here, the effective dipole moment of the molecule, $d_\mathrm{eff}$, is defined such that the long-range interaction potential is~\cite{Karman2025} 
\begin{equation}
    \bra{++}V_\mathrm{int}\ket{++}= \frac{d_\mathrm{eff}^2(1-3\cos^2\theta)}{4\pi\epsilon_0R^3},
\end{equation}
where $\theta$ is the angle between $\hat{\boldsymbol{R}}$ and $\boldsymbol{F}$.
For large $\Delta/\Omega$, the dipolar interaction due to the static electric field dominates, corresponding to a positive $a_\mathrm{d}$, while for small $\Delta/\Omega$ the dipolar interaction due to the microwave field dominates, producing a negative $a_\mathrm{d}$. These opposite contributions to the dipole length cancel at the compensation point, giving the molecule a net zero $d_\textrm{eff}$. A similar effect is seen for double microwave shielding~\cite{Karman2025}, where the the linear and circular polarizations contribute to positive and negative $a_\mathrm{d}$, respectively, and the compensation point is found at a certain ratio of their Rabi frequencies. The behaviour of the s-wave scattering length can be understood from the depth of the potential well shown in Fig.~\ref{fig:cc}(b). It is positive where the well depth is small, around the compensation point, but becomes negative for both large and small $\Delta/\Omega$ when the well depth is sufficiently large. The interactions between the molecules can therefore be tuned over a wide range, from a repulsive potential with no long-range dipolar part, to a strongly dipolar gas with positive or negative $\alpha$.

Figure \ref{fig:cc}(c) shows the rate coefficients for elastic scattering and total loss as a function of $\Delta/\Omega$. The loss decreases above the compensation point, giving access to a wide range of $\alpha$ for positive $a_\textrm{d}$. Below the compensation point, the loss rate increases and may limit the lifetimes of BECs with large negative $a_\textrm{d}$.

Our calculations have not included electron and nuclear spins. However, previous work has considered the influence of hyperfine structure on static-field shielding \cite{Mukherjee2023, Mukherjee2025b}, showing that it has little effect at most electric fields. At the strong fields considered in this paper, the rotational angular momentum is decoupled from the spins, so the hyperfine interaction will have little influence on the results presented here.

\section{Conclusions}

We have shown how a microwave field can be used to tune the interactions between static-field-shielded molecules. The microwave field introduces a dipole-dipole interaction whose strength can be tuned and whose sign is opposite to that of the static electric field. At a specific value of $\Delta/\Omega$ -- the compensation point -- the dipole length passes through zero and the scattering length is positive. Here, the rate coefficient for elastic scattering is relatively insensitive to $\Omega$, while the rate coefficient for loss falls dramatically with increasing $\Omega$. In the regime of large $\Omega$, we can expect the BEC to have a long lifetime. Tuning away from the compensation point tunes both the s-wave scattering length and the dipole length over wide ranges while maintaining low collisional loss rates. For example, when $\Omega=60$~MHz for CaF, tuning $\Delta/\Omega$ from 1.0 to 3.0 tunes $a_\textrm{d}$ from large negative to large positive values, and tunes $\alpha$ from positive to negative. Throughout this region the rate coefficient for loss is below $6 \times 10^{-13}$~cm$^3$/s and the ratio of elastic to inelastic rates is greater than 30. This large ratio means that the lifetime will be much longer than the thermalization timescale, allowing the study of the equilibrium phases of this quantum system, and of the dynamics that bring the system towards equilibrium. This ability to tune the interactions over such a wide range is crucial for exploring the novel, strongly correlated quantum phases that are expected to emerge in dipolar molecular gases. While our work has focussed on CaF, which requires large static electric fields and large Rabi frequencies, a similar approach is likely to be effective for other ultracold polar molecules including the alkali dimers.

\begin{acknowledgements}

This work has been supported by EPSRC through grants EP/W00299X/1, EP/V011499/1, EP/V011677/1 and UKRI2226.

\end{acknowledgements}

The data that support the findings of this article are openly available~\cite{zenodoTuningInteractions2026}.

\bibliography{references}

\end{document}